\documentclass[aip,amsmath,amssymb, twocolumn, 10pt, times]{revtex4-1}
\usepackage{graphicx,graphics}
\usepackage{dcolumn}
\usepackage{bm}
\usepackage{eqnarray,amsmath}
\usepackage{mathptmx}
\usepackage{etoolbox}
\usepackage{verbatim}
\usepackage{color}
\usepackage{ulem}
\usepackage[table]{xcolor}
\usepackage{booktabs}
\usepackage{float}
\usepackage[colorlinks=true]{hyperref}
\makeatletter
\def\@email#1#2{%
\endgroup
\patchcmd{\titleblock@produce} {\frontmatter@RRAPformat}
{\frontmatter@RRAPformat{\produce@RRAP{*#1\href{mailto:#2}{#2}}}\frontmatter@RRAPformat}
  {}{}
}%

\makeatother
\begin{document}

\title[]{Periodic transitions of topological charge in skyrmions confined within FeGe and Co/Pt nanodisks}

\author{R. L. Silva}
\email{ricardo.l.silva@ufes.br}
\affiliation{Departamento de Ci\^{e}ncias Naturais, Universidade Federal do Esp\'{i}rito Santo, Rodovia Governador M\'{a}rio Covas, Km 60, S\~{a}o Mateus, ES, 29932-540, Brazil.}

\author{R. C. Silva}
\email{rodrigo.c.silva@ufes.br}
\affiliation{Departamento de Ci\^{e}ncias Naturais, Universidade Federal do Esp\'{i}rito Santo, Rodovia Governador M\'{a}rio Covas, Km 60, S\~{a}o Mateus, ES, 29932-540, Brazil.}

\date{\today}

\begin{abstract}
The dynamic control of skyrmion properties such as polarity, vorticity, and topological charge is crucial for their implementation in spintronic applications. In this work, we investigate the periodic inversion of the topological charge in two distinct systems: FeGe, a bulk chiral magnet, and Co/Pt, an interfacial system with strong Dzyaloshinskii-Moriya interaction. By applying an oscillating magnetic field perpendicular to the film plane, we induce cyclic transitions in the spin texture. In FeGe, the skyrmion evolves through a $Q=1 \rightarrow 0 \rightarrow -1$ sequence via an intermediate skyrmionium state. In Co/Pt, the process involves skyrmion annihilation and re-nucleation, resulting in alternating topological charges. These results reveal distinct dynamic mechanisms for topological charge modulation, offering potential pathways for the energy-efficient control of skyrmion-based devices.
\end{abstract}

\maketitle
\section{Introduction}

Magnetic skyrmions, characterized by their nontrivial topological spin configurations, have emerged as promising candidates for next-generation data storage technologies due to their nanoscale dimensions, topological protection, and efficient current-driven mobility~\cite{Nagaosa,Cros,Cros2,Pfau,Xu}. These features render skyrmions particularly attractive for high-density, energy-efficient, and non-volatile memory applications. Their formation results from the competition between symmetric Heisenberg exchange and antisymmetric Dzyaloshinskii-Moriya interactions (DMI)~\cite{Binz,Nagaosa2,Nagaosa3}.

A key attribute of skyrmions is their quantized topological charge, defined as~\cite{Zhou1,Tokura1,Bo1}:

\begin{equation}\label{q1}
 Q=\frac{1}{4\pi} \int \int \vec{m}(\vec{r}) \cdot \left( \frac{\partial \vec{m}(\vec{r})}{\partial x} \times \frac{\partial \vec{m}(\vec{r})}{\partial y} \right) \, dx\,dy,
\end{equation}
where $\vec{m}(\vec{r})$ is the unit magnetization vector. This topological invariant provides robustness against continuous deformations that would otherwise lead to a topologically trivial (uniformly magnetized) state.

\begin{figure}[hbt]
    \centering
    \includegraphics[width=85.0mm]{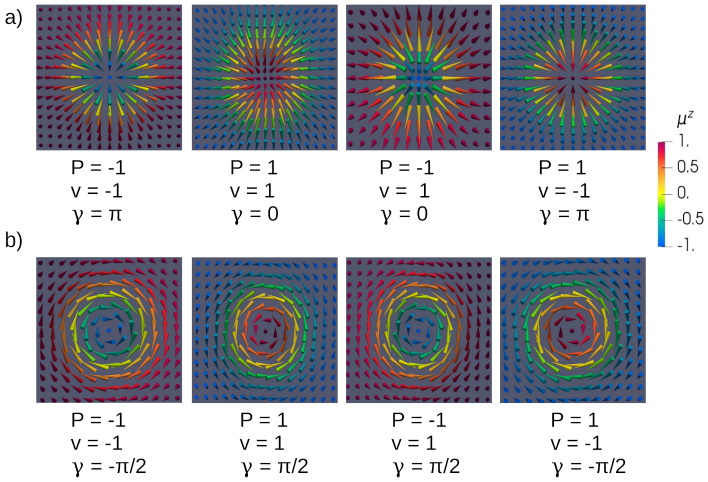}
    \caption{(Color online) Schematic illustrations of a) N\'{e}el-type and b) Bloch-type skyrmions, with representative combinations of polarity and helicity.}
    \label{fig1}
\end{figure}

For axially symmetric skyrmions, the magnetization field can be expressed as
\[
\vec{m}(\vec{r}) = (\sin\theta\cos\phi, \sin\theta\sin\phi, \cos\theta),
\]
with the spatial coordinates given by
\[
\vec{r} = (\rho\cos\varphi, \rho\sin\varphi, z),
\]
where $\theta(\rho)$ is the polar angle as a function of radial distance, and $\phi(\varphi) = v\varphi + \gamma$ is the azimuthal angle modulated by the vorticity $v$ and helicity $\gamma$. Under these conditions, the topological charge simplifies to:

\begin{equation}
 Q = -\frac{1}{4\pi}[\cos\theta]_{\rho=0}^{\rho=\infty} \cdot [\phi]_{\varphi=0}^{2\pi} = P \cdot v,
\end{equation}

where $P = -\frac{1}{2}[\cos\theta]_{\rho=0}^{\rho=\infty} = \pm 1$ defines the polarity, and $v = \frac{1}{2\pi}[\phi]_{\varphi=0}^{2\pi}$ is the in-plane vorticity. The helicity $\gamma$ does not affect the topological charge but determines the configuration of the in-plane magnetization. For $\gamma = 0$ or $\pi$, the spins point radially inward or outward, forming N\'{e}el-type skyrmions typically stabilized by interfacial DMI (see Fig.~\ref{fig1}a). For $\gamma = \pm\pi/2$, the in-plane components form an eddy-like spin texture, resulting in Bloch-type skyrmions, commonly observed in bulk chiral magnets or dipole-stabilized systems (see Fig.~\ref{fig1}b).

Independent control over polarity and vorticity is of particular interest for spintronic applications such as logic gates and multi-state memory devices. A binary change in either quantity enables the encoding of two distinct logical states. Micromagnetic simulations have shown that polarity and vorticity can be manipulated via magnetic fields, electric fields, or spin-polarized currents~\cite{Bin1,Liu1,Heo1,Bo2,Liu2}. Experimentally, such transitions have been observed in FeGe nanodisks and MnNiGa thin films under applied magnetic fields~\cite{Li1,Yao1}. In FeGe, polarity and vorticity invert simultaneously, while in MnNiGa, only the vorticity changes, with the polarity remaining unchanged.

To probe these dynamic transitions, techniques such as polarized resonant elastic X-ray scattering (REX)~\cite{Yao1, Laan1, Laan2, Laan3, Hoffman} have been employed. In REX, magnetization-dependent scattering of X-rays encodes information about the spin texture, which is captured by a photodiode point detector capable of resolving intensity changes on nanosecond timescales. This technique enables real-time observation of skyrmion dynamics, including rapid topological transitions.

In this work, we investigate the dynamical inversion of skyrmion polarity and vorticity, which occurs simultaneously, resulting in a full inversion of the topological charge, from \( Q=+1 \) to \( Q=-1 \). Our results demonstrate that applying an oscillating out-of-plane magnetic field enables deterministic and periodic switching of the topological state in both FeGe and Co/Pt nanodisks. In FeGe, the topological charge undergoes a cyclic transition through an intermediate skyrmionium state, from a skyrmion with $Q=1$ to a skyrmionium ($Q=0$), and then to an inverted skyrmion with $Q=-1$. In Co/Pt, the transition involves annihilation into a uniformly magnetized monodomain, followed by the nucleation of a new skyrmion with reversed polarity and vorticity. These findings reveal the existence of a robust, field-driven mechanism for periodic topological inversion, providing reversible control of distinct spin textures in confined magnetic geometries.

\section{Methods}
To model our systems, we use a Hamiltonian:
\begin{equation}\label{eq:Hamiltonian}
H = H_{\mathrm{Exc}} + H_{\mathrm{DMI}} + H_{\mathrm{Ani}} + H_{\mathrm{Zeeman}},
\end{equation}
where
\begin{eqnarray}
  H_{\mathrm{Exc}} &=& -J_{ex}\sum_{\langle i,j \rangle} \left(\vec{\mu}_{i} \cdot \vec{\mu}_{j}\right)\,,\nonumber\\
  H_{\mathrm{DMI}} &=& -\sum_{\langle i,j \rangle}\vec{D}_{ij}\cdot\left(\vec{\mu}_{i}\times \vec{\mu}_{j}\right)\,,\nonumber \\
  H_{\mathrm{Ani}} &=& -K_{eff} \sum_{i}(\mu_{i}^z)^2\,,\nonumber \\
  H_{\mathrm{Zeeman}} &=& -M_{s}a^{3}\sum_{i} \vec{H}_{ext} \cdot \vec{\mu}_{i} \,.\nonumber
\end{eqnarray}
which is actually an extended two-dimensional Heisenberg model. Here, the spin texture is represented by an array of unit-norm dimensionless vectors with the form $\vec{\mu}_{i} = (\mu_{i}^{x}, \mu_{i}^{y}, \mu_{i}^{z})$, which represents the magnetic moment at the lattice site $i$. The ferromagnetic exchange interaction between two nearest neighboring spins is given by $J_{ex} = 2aA_{ex}$, where $A_{ex}$ is the exchange stiffness and $a$ is the lattice constant. The summation $\langle i,j \rangle$ runs over nearest neighbors. The second term describes the Dzyaloshinskii-Moriya interaction (DMI), with $|\vec{D}_{ij}| = D$ denoting the magnitude of the DMI vector. The third term represents uniaxial anisotropy along the $z$-axis, with $K_{eff}$ being the effective anisotropy constant. The fourth term is the Zeeman interaction, where an external oscillating magnetic field along the \textit{z}-direction is given by $\vec{H}_{ext} = h_{0z} \sin(2\pi f_{0}t)\, \hat{z}$. In this study, dipolar interactions are effectively incorporated into the anisotropy term with the strength  $K_{eff} = K - \frac{1}{2}\mu_{0}M_{s}^{2}$ in Eq.~(\ref{eq:Hamiltonian}).

The initial skyrmion profile is constructed using the analytical ansatz proposed in Ref.~\cite{Silva}:
\begin{eqnarray}\label{eq:skyrmion_profile}
\vec{\mu}_{ij} &=& \left( \sin\theta(r)\cos\phi,\ \sin\theta(r)\sin\phi,\ \cos\theta(r) \right), \nonumber \\
\\
\mathrm{with} \quad
r &=& \sqrt{(i - i_0)^2 + (j - j_0)^2}, \nonumber \\
\theta(r) &=& 2\arctan\left( \frac{r}{R_s} \right), \nonumber \\
\phi &=& \arctan\left( \frac{j - j_0}{i - i_0} \right), \nonumber
\end{eqnarray}
where \( (i_0, j_0) \) denotes the coordination of the skyrmion center, and  \( R_s \) represents its radius.

In this equation, the angle \( \phi \) denotes the in-plane azimuthal angle with respect to the skyrmion center. The spin configuration described by Eq.~(\ref{eq:skyrmion_profile}) corresponds to a \textit{N\'{e}el-type} skyrmion, where the in-plane magnetization vectors point radially inward or outward. A \textit{Bloch-type} skyrmion is obtained by shifting the azimuthal angle by \( \pi/2 \), i.e., $\phi \rightarrow \phi + \pi/2$, leading to an eddy-like in-plane spin texture around the skyrmion core.

To simulate the magnetization dynamics, we numerically solve the Landau-Lifshitz-Gilbert (LLG) equation:
\begin{equation}\label{eq:LLG}
    \frac{\partial\vec{\mu}_{i}}{\partial \tau} = -\frac{1}{1+\alpha^{2}}\left[\vec{\mu}_{i}\times
\vec{b}_{i}+\alpha \vec{\mu}_{i}\times \left(\vec{\mu}_{i} \times \vec{b}_{i}\right)\right],
\end{equation}
where $\alpha$ is the Gilbert damping constant and $\vec{b}_{i} = -\frac{1}{J_{ex}}\frac{\partial \mathcal{H}}{\partial \vec{\mu}_{i}}$ is the dimensionless local effective field at site $i$. The dimensionless time $\tau$ is related to real time through $dt = \nu d\tau$, where $\nu = \frac{M_{s}a^{3}}{\gamma J}$ and $\gamma$ is the electronic gyromagnetic ratio. Time integration is performed using a fourth-order Runge-Kutta scheme with a time step $\delta \tau = 0.001$, implemented in a custom-developed code.

Simulations are carried out for the nanodisks consisting of FeGe and Co/Pt with the material parameters listed in Table~\ref{tab:table1}.

\begin{table}[hbt]
    \begin{center}
        \caption{Magnetic parameters used in the simulations~\cite{Saavedra,Sampaio}.}
        \label{tab:table1}
        \begin{tabular}{c|c|c}
      \hline
      \hline
      Parameters              & FeGe             & Co/Pt            \\
      \hline
      $A_{ex}$                & 8.78 pJ/m        & 15 pJ/m          \\
      $D$                     & 1.58 mJ/m$^2$    & 4.0 mJ/m$^2$     \\
      $K$                     & 0.1 MJ/m$^3$     & 0.8 MJ/m$^3$     \\
      $M_{s}$                 & 0.385 MA/m       & 0.58 MA/m        \\
      $\lambda_{ex}$         & 9.7 nm           & 8.4 nm           \\
      $\alpha$               & 0.3              & 0.3              \\
      \hline
      \hline
        \end{tabular}
    \end{center}
\end{table}

\begin{figure}[hbt]
    \centering
    \includegraphics[width=85.0mm]{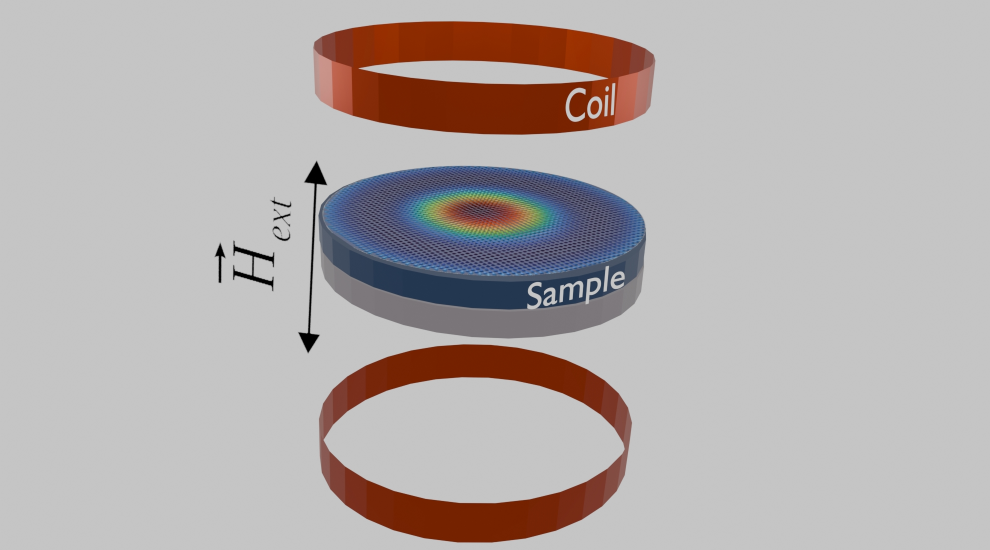}
    \caption{Diagram of the experimental setup: A nanodisk hosting a magnetic skyrmion is subjected to an oscillating magnetic field $\vec{H}_{\mathrm{ext}}$, generated by a nearby Radio Frequency coil (RF coil).}
    \label{fig2}
\end{figure}

We adopt a lattice constant of $a = 2.5$ nm for both systems. The nanodisks have a radius of 150 nm and a thickness of 10 nm, which are in agreement with experimentally relevant geometries for FeGe and Co/Pt thin films~\cite{Law,Wang,Anton,Tai}. Figure~\ref{fig2} presents a schematic representation of the setup, where a nanodisk hosting a skyrmion is subjected to an oscillating magnetic field generated by a Radio Frequency coil ( RF coil).

To analyze the spin dynamics, we compute the neutron scattering function $S(\vec{q},\omega)$, which captures information about both momentum transfer $\vec{q}$ and energy transfer $\omega$. This function serves as a key tool to probe spin excitations in confined  topologically non-trivial magnetic systems. It is defined as~\cite{ABunker}:
\begin{eqnarray}\label{eq:NSF}
    S^k(\vec{q},\omega ) &=& \sum_{\vec{r},\vec{r'}} e^{i\vec{q}\cdot(\vec{r}-\vec{r'})}
    \int_{-\infty}^\infty e^{i\omega t} C^{k}(\vec{r}-\vec{r'},t) \frac{dt}{2\pi},
\end{eqnarray}
where $C^k(\vec{r}-\vec{r'},t)$ is the space- and time-displaced spin-spin correlation function:
\begin{equation}\label{eq:correlation}
 C^k(\vec{r}-\vec{r'},t)= \langle \mu_{\vec{r}}^k(t) \mu_{\vec{r'}}^k(0) \rangle - \langle \mu_{\vec{r}}^k(t)\rangle \langle \mu_{\vec{r'}}^k(0) \rangle,
\end{equation}
with $k = x, y,$ or $z$, and where the positions $\vec{r}$ and $\vec{r'}$ are expressed in units of the lattice constant $a$. In the ferromagnetic regime, $\mu_{\vec{r}}^k(t)$ denotes the $k$-component of the spin at site $\vec{r}$ and time $t$.

We focus on $S^{z}(\vec{q},\omega)$ to identify characteristic skyrmion excitation modes in the nanodisk geometry. 
These include low-frequency breathing modes, corresponding to radial oscillations of the spin texture, and higher-frequency azimuthal modes, characterized by angular modulations and rotational dynamics of the skyrmion. We also observe localized and hybridized internal modes resulting from the interplay of confinement, topology, and boundary conditions~\cite{Guslienko,Vigo,Bin,Fert,Lake}.

To evaluate the topological charge $Q$ of the evolving spin texture, we employ a discretized scheme based on the solid angle subtended by neighboring spins. Eq.~(\ref{q1}) is valid only for continuous spin textures and thus is not applicable to discrete spin systems. Instead, we use the geometrical method proposed by Berg and L\"{u}scher~\cite{BergLuscher1981}, where each square plaquette is divided into two triangles, and the solid angle $\Omega$ subtended by the three spins $\vec{a}$, $\vec{b}$, and $\vec{c}$ of the one triangle is computed with:

\begin{equation}
\Omega(\vec{a}, \vec{b}, \vec{c}) = 2 \arctan \left( \frac{ \vec{a} \cdot (\vec{b} \times \vec{c}) }{1 + \vec{a} \cdot \vec{b} + \vec{b} \cdot \vec{c} + \vec{c} \cdot \vec{a}} \right).
\end{equation}

This formulation is numerically stable and invariant under global spin rotations. The topological charge of a skyrmion is calculated by summing over all triangles in the skyrmion. This discrete formulation is especially suitable for atomistic spin simulations, allowing for real-time monitoring of topological transitions such as skyrmion-antiskyrmion conversion, annihilation events, and the transient formation of skyrmionium states under time-dependent external stimuli.

\section{Results and Discussion}
\subsection{Ground State}
To obtain the skyrmionic configurations in FeGe and Co/Pt nanodisks, we relax an initial spin configuration obtained from Eq.~(\ref{eq:skyrmion_profile}), by minimizing the total energy defined by Eq.~(\ref{eq:Hamiltonian}). For FeGe nanodisk, we employ a Bloch-type DMI that acts both in-plane and out-of-plane directions, consistent with its bulk chiral nature. The discrete form of the Bloch DMI implemented in our simulation code is:
\begin{eqnarray}
 \mathcal{H}_{\mathrm{DMI}}^{\mathrm{Bloch}} &=& D \sum_i \left( \vec{\mu}_i \times \vec{\mu}_{i+\hat{x}} \cdot \hat{y} - \vec{\mu}_i \times \vec{\mu}_{i+\hat{y}} \cdot \hat{x} \right. \nonumber \\
 && \left. + \vec{\mu}_i \times \vec{\mu}_{i+\hat{z}} \cdot \hat{x} - \vec{\mu}_i \times \vec{\mu}_{i+\hat{x}} \cdot \hat{z} \right),
\end{eqnarray}
which promotes helical twisting of spins in all spatial directions.

For the Co/Pt system, where DMI originates from strong interfacial spin-orbit coupling, we consider a N\'{e}el-type DMI restricted to the plane:
\begin{eqnarray}
 \mathcal{H}_{\mathrm{DMI}}^{\mathrm{N\acute{e}el}} = D \sum_i \left( \vec{\mu}_i \times \vec{\mu}_{i+\hat{x}} \cdot \hat{x} + \vec{\mu}_i \times \vec{\mu}_{i+\hat{y}} \cdot \hat{y} \right),
\end{eqnarray}
which stabilizes radially symmetric textures typical of interfacial skyrmions.

\begin{figure}[hbt]
    \centering
    \includegraphics[width=85.0mm]{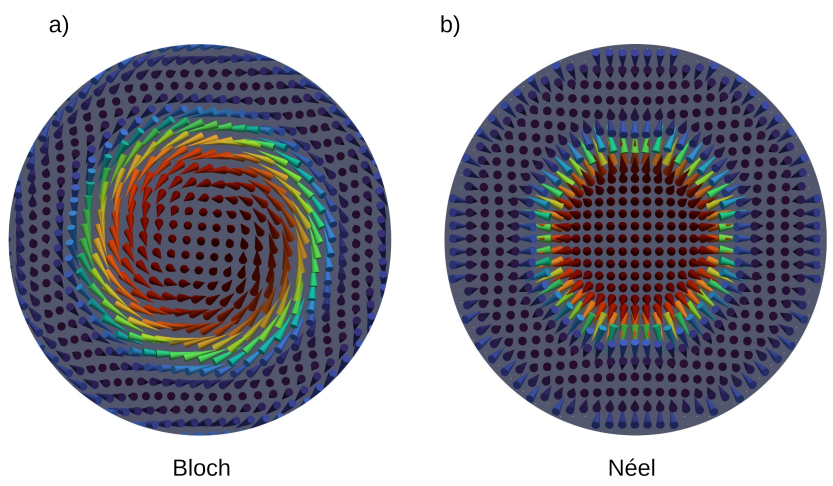}
    \caption{Ground-state spin textures in nanodisks. (a) Bloch-type skyrmion stabilized in a FeGe nanodisk due to bulk DMI. (b) N\'{e}el-type skyrmion stabilized in a Co/Pt nanodisk due to interfacial DMI.}
    \label{grd}
\end{figure}

The spin relaxation is performed by numerically integrating the LLG equation with a high damping constant ($\alpha = 1$) using a fourth-order Runge-Kutta method. Figure~\ref{grd} (a) and (b) show the ground-state spin textures for the FeGe and Co/Pt nanodisks, respectively. In both cases, the relaxed skyrmions exhibit a radius of approximately 40 nm.

\subsection{Dynamic Modes of FeGe and Co/Pt Nanodisks}

To probe the dynamic response of skyrmions to external excitations, we compute the dynamic structure factor $S^{z}(\vec{q},2\pi f)$ under a pulsed magnetic field applied along the $z$-axis. The pulse has a sinc-shaped profile:
\[
H_{\mathrm{sinc}}(t) = h_0 \, \frac{\sin\left[2\pi f_{\mathrm{sinc}} (t - t_0)\right]}{2\pi f_{\mathrm{sinc}} (t - t_0)},
\]
where $h_0$ is the amplitude, $f_{\mathrm{sinc}}$ is the central frequency, and $t_0$ is the pulse center. For FeGe, the pulse has $h_0 = 5$ mT and $f_{\mathrm{sinc}} = 3.4$ GHz; for Co/Pt, $h_0 = 4$ mT and $f_{\mathrm{sinc}} = 6$ GHz. In both cases, $t_0 = 10$ ns and the total simulation time is 20 ns.

\begin{figure}[hbt]
    \centering
    \includegraphics[width=85.0mm]{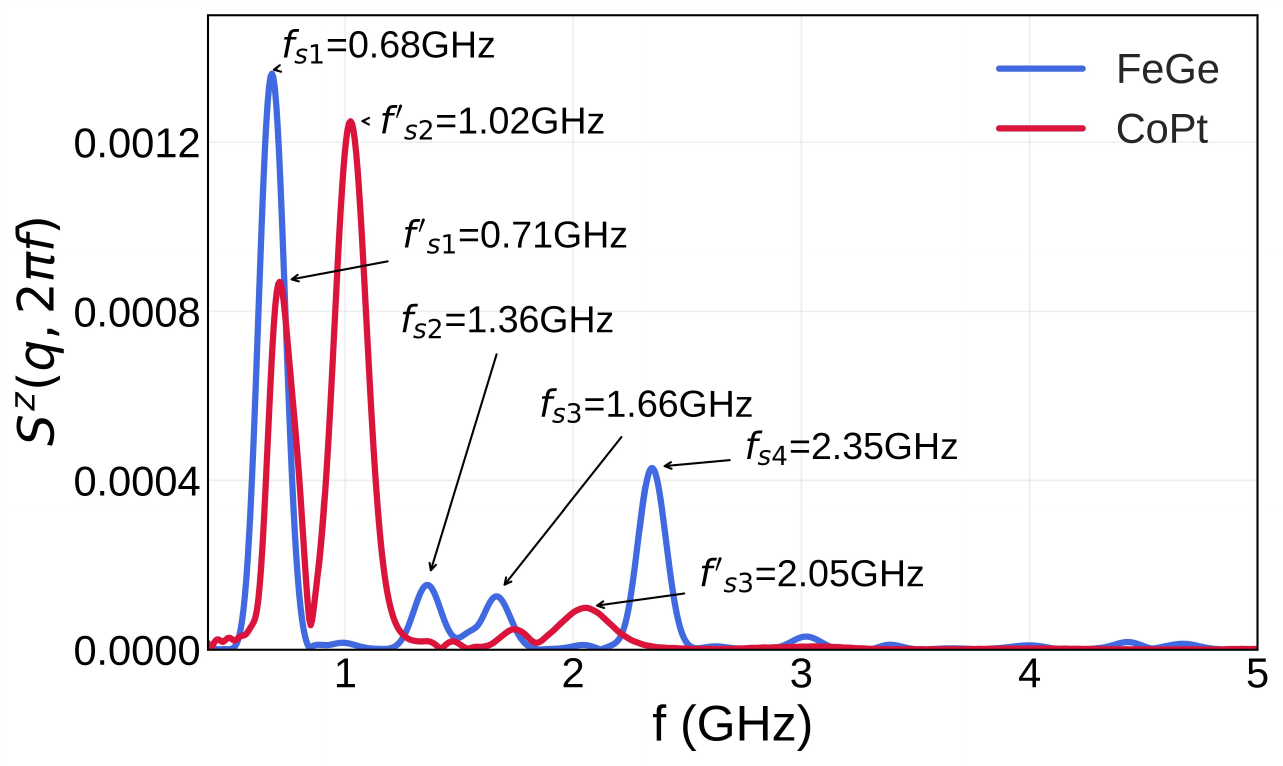}
    \caption{Dynamic structure factor $S^{z}(\vec{q},2\pi f)$ as a function of frequency for magnetic skyrmions in FeGe (blue) and Co/Pt (red) nanodisks, obtained from atomistic simulations. Resonance peaks $f_{si}$ and $f'_{si}$ correspond to intrinsic skyrmion modes.}
    \label{fig3}
\end{figure}

The dynamic structure factor is evaluated at a wavevector with modulus $q = 2\pi / R$, where $R$ is the nanodisk radius, corresponding to collective excitations spanning the entire structure. Figure~\ref{fig3} shows the resulting $S^{z}(\vec{q},2\pi f)$ spectra for both materials, revealing distinct internal modes excited by the sinc-shaped magnetic field.

\begin{figure}[hbt]
    \centering
    \includegraphics[width=85.0mm]{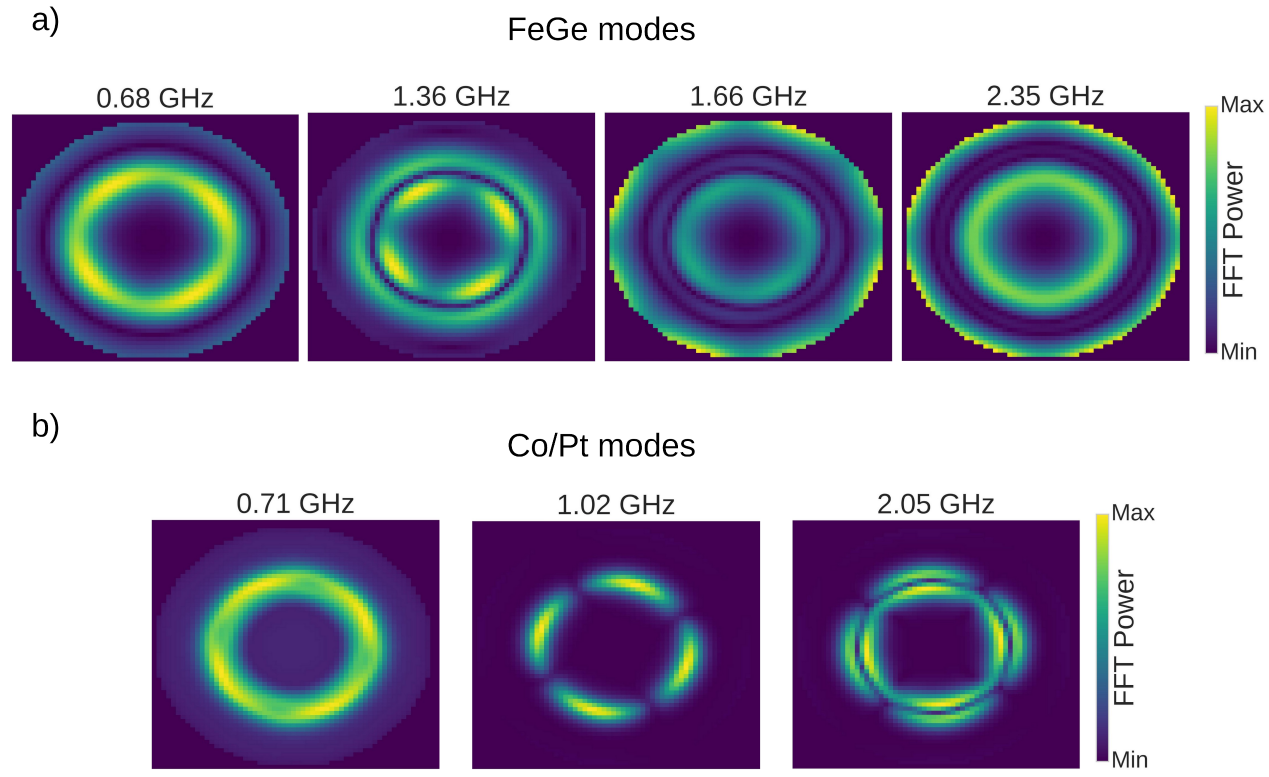}
    \caption{Unrolled plane-view maps of FFT power distributions at selected resonance frequencies for (a) FeGe and (b) Co/Pt nanodisks.}
    \label{md}
\end{figure}

Figure~\ref{md} presents the spatial maps of the dynamic mode profiles at selected resonance frequencies. At low frequencies - 0.68 GHz (FeGe) and 0.71 GHz (Co/Pt) - the modes exhibit radial symmetry consistent with the breathing mode. As the frequency increases (1.36 GHz and 1.02 GHz), angular modulations appear, indicating rotational or azimuthal core modes. At even higher frequencies (1.66 and 2.35 GHz in FeGe; 2.05 GHz in Co/Pt), the modes display complex structures with concentric and angular nodes, showing typical features of higher-order confined spin-wave modes.

\subsection{Periodic Topological Charge Switching}
\begin{figure*}[hbt]
    \centering
    \includegraphics[width=170.0mm]{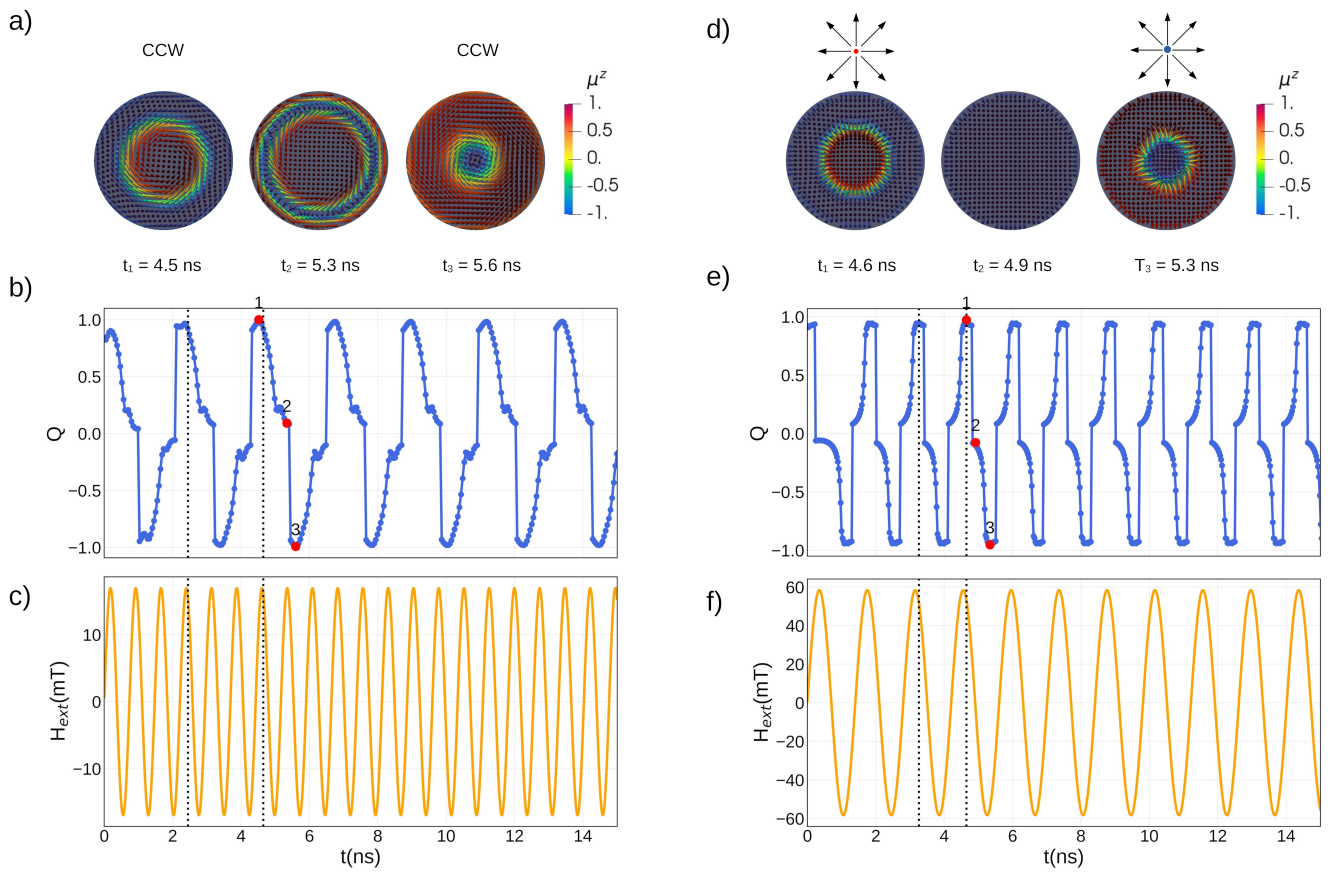}
    \caption{(a) and (d) Snapshots of the spin texture under oscillating out-of-plane magnetic fields, highlighting the dynamical evolution of the skyrmion structure for FeGe and Co/Pt nanodisks. The temporal evolution of the topological charge $Q$ is shown in (b) and (e), while the temporal dependence of the external field $H_{ext}$ is presented in (c) and (f). Selected points corresponding to the textures in (a) and (d) are indicated by red markers. Periodic inversion of the skyrmion core magnetization leads to dynamic switching of the topological charge between $+1$, $0$ an $-1$, without disruption of the continuous spin texture or formation of topological defects. The results demonstrate reversible control of the skyrmion polarity and topological charge through external field excitation.}
    \label{sw}
\end{figure*}

 After determining the eigenfrequencies of the skyrmion stabilized in the nanodisk, we investigate the topological charge switching by applying an out-of-plane alternating magnetic field at resonance, $\vec{H}_{\mathrm{ext}} = h_{0z} \sin(2\pi f_{0} t)\cdot \hat{z}$, where $h_{0z}$ is the amplitude and $f_{0}$ is the excitation frequency. We use the internal excitation modes as the frequencies of the alternating magnetic field. For modes with stronger intensities, such as \(f_{s3} = 1.66\) GHz, \(f_{s4} = 2.35\) GHz, and \(f'_{s3} = 2.05\) GHz, the field can induce a change in the skyrmion polarity within the nanodisks. However, this change is not repetitive, it is non-periodic. Periodic transitions are instead observed in weaker modes at frequencies $f_{s1}=0.68$ GHz and $f'_{s1}=0.71$ GHz, along with intermediate modes at $f_{s2}=1.36$ GHz and $f'_{s2}=1.02$ GHz. Figures~\ref{sw}(a) and \ref{sw}(d) illustrate the transition from a skyrmion to a skyrmionium and back to a skyrmion for the FeGe nanodisk, and the transition between skyrmion and monodomain-like skyrmion states for the Co/Pt nanodisk, respectively. Furthermore, the corresponding temporal evolution of the topological charge and the applied magnetic field is shown in Figs.~\ref{sw}(b), \ref{sw}(c), \ref{sw}(e), and \ref{sw}(f).

Inside the FeGe nanodisk, a skyrmion with an initially upward core, counterclockwise-swirling in-plane spin texture, and  topological charge $Q = +1$ evolves through a multi-step transition under an oscillating magnetic field. The transformation proceeds via an intermediate skyrmionium state ($Q = 0$) at $t_{2} = 5.3\,\mathrm{ns}$ before culminating in an inverted skyrmion ($Q=-1$) with a downward core at $t_{3} = 5.6\,\mathrm{ns}$  (see Fig.~\ref{sw}(a)). This sequence $Q = +1 \rightarrow 0 \rightarrow -1$ is driven by a resonant AC field ($f_{0} = 1.36 \, \mathrm{GHz}$, $h_{0z} = 17 \,\mathrm{mT}$) applied over $\Delta t = 1.1 \, \mathrm{ns}$. Remarkably, this transition exhibits a subharmonic response, with the topological charge switching at a frequency $f_{sw} = f_{0}/3$. This fractional response arises from the nonlinear dynamics of the system: the skyrmion initially exhibits breathing-mode oscillations, accumulating energy over multiple field cycles before overcoming the energy barrier to form the skyrmionium. A subsequent barrier must be traversed to collapse the skyrmionium into the final inverted state. Such multi-stage transitions are hallmarks of resonantly driven topological systems, where energy storage in internal modes (e.g., breathing modes) delays the response until critical thresholds are reached~\cite{Wang2015, Schutte2014}. Similar subharmonic behavior ($f_{sw}=f_{0}/N$, with odd $N$) has been reported for a magnetic votex systems~\cite{Gypens2022}, underscoring the universal nature of nonlinear mode coupling in confined spin textures. Thus, the observed $f_{0}/3$ ratio in FeGe is a direct consequence of fundamental nonlinear dynamics in topological spin systems, reflecting the three distinct topological stages of the transformation.

In contrast to the multi-stage transition in FeGe, the Co/Pt nanodisk exhibits an abrupt topological inversion. The initial skyrmion ($Q= +1$), characterized by upward core polarization and divergent in-plane magnetization, undergoes complete annihilation, yielding a monodomain state with uniform negative out-of-plane magnetization. Subsequently, a new skyrmion ($Q=-1$) nucleates with inverted core polarization while retaining the divergent magnetization profile (Fig.~\ref{sw}(d)). This direct $Q = +1 \rightarrow -1$ transition is triggered by an out-of-plane AC field ($f_{0}=f'_{s2}=1.02$ GHz, $h_{0z}=73$ mT), with the topological charge oscillation frequency, showed in Fig.~\ref{sw}(e) (dashed lines), locking precisely to the driving field frequency (Fig.~\ref{sw}(f)). The stark difference in transition dynamics, immediate field synchronization in Co/Pt versus the delayed $f_{0}/3$ response in FeGe, highlights the critical role of material-specific stiffness and energy landscapes in topological switching. Full dynamic sequences are available in Videos 1 (FeGe) and 2 (Co/Pt).

We applied off-resonance alternating magnetic fields to both FeGe and Co/Pt nanodisks, varying the frequency from $f_{0} = 0.15$ GHz to $f_{0} = 0.55$ GHz in increments of $\Delta f = 0.05$ GHz. For the FeGe nanodisk, we observed that the topological transition occurs without the formation of an intermediate skyrmionium state ($Q = 0$). In this process, the initial skyrmion, which has an upward core polarity and counterclockwise in-plane magnetization ($Q = +1$), is annihilated. This is followed by the formation of a monodomain with negative out-of-plane magnetization. Afterwards, a transient skyrmion with downward core polarity and tangential in-plane magnetization, characteristic of a Bloch-type configuration, is formed. This structure corresponds to a topological charge ($Q = -1$), and emerges briefly before the system continues its evolution. The frequency of the topological charge oscillation in this regime matches that of the applied AC field.

In the case of the Co/Pt nanodisk, the topological transition proceeds in the same manner as observed under resonant excitation, where the driving frequency was determined from the system's normal modes obtained via a sinc-shaped magnetic pulse. In both cases, the skyrmion undergoes periodic annihilation and re-nucleation without the formation of a skyrmionium state, and the frequency of topological charge inversion matches that of the applied alternating field, i.e., \( f_Q = f_H \).

\begin{figure}[hbt]
    \centering
    \includegraphics[width=85.0mm]{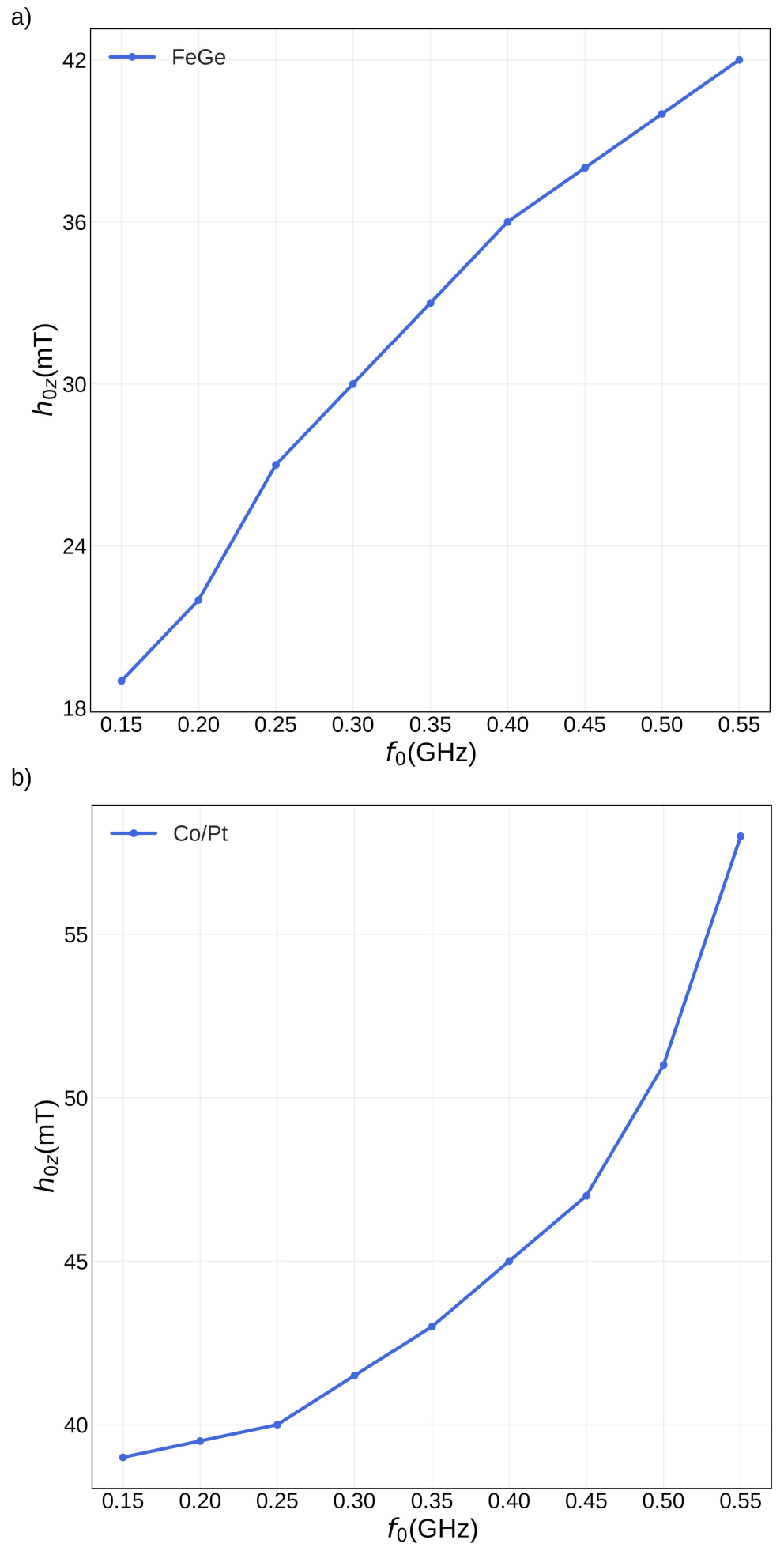}
    \caption{Minimum out-of-plane excitation field amplitude required to induce topological charge inversion as a function of frequency in magnetic nanodisks.
    (a) FeGe: the threshold field increases approximately linearly from 20 mT to 40 mT, indicating strong frequency sensitivity.
    (b) Co/Pt: the threshold varies from 40 mT to 57.5 mT, showing a sharper frequency response but a smaller relative variation, consistent with a more rigid skyrmion profile.}
    \label{fr}
\end{figure}

Figure~\ref{fr} shows the amplitude of the magnetic field as a function of frequency. This value of $h_{0z}$ corresponds to the minimum field amplitude required to induce the transitions.

In FeGe [Fig.~\ref{fr}(a)], the threshold grows almost linearly from 18.5 mT to 41.5 mT over the interval $f_0 = 0.15$ - 0.55 GHz, representing a relative increase of approximately 124\%. This gradual variation reflects the soft nature of the spin texture in FeGe, which is stabilized by bulk DMI and low magnetic anisotropy, allowing the skyrmion to couple effectively to a broad frequency spectrum. As the frequency increases, progressively higher-energy modes must be accessed to trigger the transition, which demands stronger excitation fields.

In contrast, Co/Pt [Fig.~\ref{fr}(b)] shows a steeper nonlinear dependence with a higher absolute threshold (38-58 mT), yet with a smaller relative increase of only 52\%. This behavior is consistent with a stiffer spin texture, arising from strong perpendicular magnetic anisotropy and interfacial DMI. The response to frequency is more abrupt: small variations in $f_0$ lead to considerable increases in the required field. This suggests that fewer modes are efficiently excited, and the system requires a larger field to overcome sharper energy barriers concentrated around resonant frequencies.

These results demonstrate that the distinct dynamic responses of FeGe and Co/Pt nanodisks cannot be attributed to exchange stiffness alone. Instead, they arise from a complex interplay of magnetic parameters and structural characteristics. In Co/Pt, the skyrmion profile is stiffer due to strong perpendicular magnetic anisotropy and interfacial DMI, which constrain the number of accessible excitation modes and demand higher field amplitudes for topological switching. In contrast, FeGe exhibits bulk-type DMI and low anisotropy, resulting in a softer, more deformable spin texture capable of coupling to a broader range of resonant frequencies. Dimensionality further contributes to this difference: as a bulk chiral magnet, FeGe supports three-dimensional spin twisting and a richer spectrum of internal modes, whereas Co/Pt behaves as a quasi-two-dimensional system with enhanced confinement effects. Thus, the contrasting dynamic behaviors arise from the combined influence of anisotropy, DMI symmetry, exchange stiffness, and geometrical confinement.

\section{Conclusions and Prospects}
We have shown that the periodic inversion of skyrmion topological charge can be achieved in magnetic nanodisks using resonant and near-resonant excitation with out-of-plane alternating magnetic fields. This dynamic behavior, particularly observed in FeGe and Co/Pt systems, enables controlled transitions between skyrmion states with topological charges of \( Q = +1 \), 0, and \( -1 \), without generating singular defects or causing irreversible annihilation.

In the case of FeGe, the charge oscillation includes intermediate skyrmionium states, highlighting a smooth and continuous transformation of the spin texture. In contrast, Co/Pt displays direct and abrupt switching between skyrmion and monodomain-like states, reflecting its higher magnetic stiffness.

Most notably, we identified specific excitation frequencies that result in robust and periodic topological switching. In this scenario, the charge oscillates over time with a well-defined frequency, which may either match or be detuned from the driving field. This reversibility emphasizes the potential for dynamic encoding information in skyrmion topology, with minimal energy dissipation and without permanent structural changes.

These findings establish a new operational regime for skyrmionic systems under time-dependent fields and suggest promising applications for topological frequency-encoded memory.

\section*{Acknowledgements}
The authors thank CNPq and FAPES (Brazilian agencies) for financial support.

\end{document}